# Operational and biomechanical evaluation of a wrist exoskeleton prototype for assisting meat cutting tasks

Aurélie Tomezzoli, Mathieu Gréau, and Charles Pontonnier

*Abstract*—Although a growing number of exoskeletons have been developed for occupational applications, wrist exoskeletons remain relatively rare. However, in the meat processing industry, elbow and hand-wrist musculoskeletal disorders are particularly common. The aim of this study was to assess the potential effectiveness and risks of a 670g wrist exoskeleton prototype designed to assist operators during meat cutting tasks. Six professional butchers performed three standardized tasks reproducing meat cutting gestures in foam, in three randomized experimental conditions: (1) without exoskeleton, (2) wearing the exoskeleton passive, with brakes off and (3) using it with brakes activated, locked in a static position. Cutting forces were recorded using an instrumented table, joint angles using an optoelectronic motion capture system, muscle activity using surface EMG and user experience was assessed using questionnaires. The cutting inaccuracy was defined as the area between the prescribed task and the actual cut on the foam surface. Joint torques were estimated by inverse dynamics with and without taking the exoskeleton's mass into account, to isolate its effect. Linear mixed-effects statistical models were fitted. With the exoskeleton active during tasks, EMG activity was decreased of up to 18.7% (p<0.01 to p<0.001) in the wrist flexors and increased of up to 61.7% (not significant to p<0.05) in the upper trapezius. Shoulder elevation joint torques were increased of up to 39.7% (p<0.001), mainly due to the exoskeleton mass. The proposed multi-criteria exoskeleton evaluation has provided guidance for following prototyping stages. Too heavy wrist exoskeletons could increase the risk of shoulder tendinitis for such tasks.

## I. INTRODUCTION

PROVIDING physical assistance to operators in their work is a fast-growing approach that complements traditional ergonomics interventions to prevent the onset or recurrences of musculoskeletal disorders (MSDs) in the workplace. Cobots [1] or exoskeletons are therefore currently being developed [2], [3]. For upper limbs, recent literature reviews highlight the existence of a large variety of exoskeletons assisting shoulder in tasks such as lifting, carrying or performing overhead work [2], [4], [5], [6] while few have been developed for the hand and wrist [7], [8].

In the meat processing industry, work-related MSDs are a major health issue, with a prevalence 10 to 15 times higher than in the general population [9]. Elbow and hand-wrist MSDs are over-represented, even though neck and shoulder MSDs prevalence remains high [10]. Biomechanical characterization of pork-cutting tasks performed by a limited number of butchers showed that the estimated wrist flexion and ulnar or radial deviation torques were unsustainable for most operators [11]. These considerations suggest to mechanically assist butchers in meat cutting tasks [12], [13], more precisely to mitigate the exposure to forceful exertions at the wrist. Assistance provided by an exoskeleton is expected to retains the dexterity and versatility required for meat cutting tasks [12] while reducing the constraints undergone, and can therefore be considered for this application. Furthermore, passive assistance of operators in radial or ulnar deviation would overload them in the opposite direction, suggesting that assistance in this degree of freedom (dof) must be active.

The design of wrist exoskeletons is complex, partly due to anatomical factors. One part of a wrist exoskeleton is indeed fixed on the hand, the other on the forearm. The hand is a mobile, multi-joint body part [14]. The forearm is mobile too, as the pronosupination movement involves rotation around its longitudinal axis [15], [16]. Any loss of wrist mobility could hinder the achievement of the task to be performed or lead to kinematic adaptations of proximal joints i.e., at elbow, shoulder, and spine, which could increase the physical load at these levels. Moreover, as in the case of wearing a medical splint or cast, the pressure of the exoskeleton on the upper limb can cause discomfort or damage the skin or nerves [17]. In addition to the problems inherent to this body region anatomy, the hand tactile function must also be preserved when fine motor control is required. The volume of the exoskeleton can be a hindrance to perform the task. The exoskeleton can also be damaged by objects being handled. Finally, in the event of contact with food, microbiological and chemical risks must be considered.

This study has been funded by the Région Bretagne, within the frame of the EXOSCARNE 2.0 project, and supported by French government funding managed by the National Research Agency under the Investments for the Future program (PIA) with the grant ANR-21-ESRE-0030 (CONTINUUM project). *(Corresponding author: Aurélie Tomezzoli).*

Aurélie Tomezzoli and Charles Pontonnier are with the Univ Rennes, CNRS, Inria, IRISA - UMR 6074, F-35000 Rennes, France. Mathieu Gréau is with Lab4i, 35830 Betton, France (e-mail: aurelie.tomezzoli@ens-rennes.fr, mgreau@lab4i.io, charles.pontonnier@ens-rennes.fr).

The study was approved by a national ethics committee (Comité Opérationnel d'Evaluation des Risques Légaux et Ethiques, COERLE, n° 2023-25). Each participant signed an informed consent form prior to participating in the experiment.

The questionnaire developed and used in this study is available online in English and French at http://ieeexplore.ieee.org, as are the supplementary figures mentioned in the main text.

To assess an exoskeleton prototype, a combination of evaluation criteria related to health objectives and industrial constraints can therefore be proposed [18]. MSDs are multifactorial diseases in which chronic hyper-solicitation of soft tissues plays a major role. They may only appear after prolonged exposure. At the prototype evaluation stage, intermediate criteria that enable quick and analytical risk assessment are sought. Risk factors for MSDs fulfill this role. Physical risk factors are primarily awkward postures or forceful exertions in the context of repetitive or prolonged tasks [10], [19–23]. These factors can be studied using biomechanical quantities such as joint angles, joint torques and electromyographic (EMG) activities [24]. The exoskeleton must also be compatible with the targeted industrial activity and must therefore be usable, as defined for example by the ISO standard 9241-11 [25]. In the meat processing industry, neither the duration of task completion, nor the cutting accuracy must be affected. Operator satisfaction is also key to ensure the exoskeleton acceptance and appropriation [26], [27].

The aim of this study was to assess, early in the development process, the potential effectiveness and risks of a prototype of wrist exoskeleton intended for long-term daily use in an industrial environment, based on both biomechanical variables related to the risk of MSDs and on usability criteria, during standardized tasks reproducing meat cutting gestures in foam. The hypothesis was that the exoskeleton was effective in reducing physical exertion, particularly in wrist flexion, and was usable. Concurrently, we also aimed to identify possible causes of physical overload related to the exoskeleton design.

## II. METHODS

### A. Data collection

Six butchers (45.5±7.9 years, 23.0±11.7 years of experience, height 175.5±4.6 cm, weight 80.1±14.5 kg), right-handed, with no occupational disease declared in the previous 6 months, were recruited. The equipment for cutting was a 110mm knife commonly used in participants' factory, and a chainmail apron, which is the personal protective equipment required when using butcher knives.

The exoskeleton weighed 670g. A modified semi-rigid medical splint was fixed to the forearm with two Velcro straps, to optimize comfort and power transmission. A two millimeters neoprene glove covered the thumb and all interdigital spaces, providing a flexible, slip-reducing element on the hand, which minimized both interference with the grip and loss of tactile perception (Fig. 1.A). The actuation of the system consisted in two bars articulated with spherical joints on the glove and forearm, associated with each dof of the exoskeleton (radio-ulnar deviation and wrist flexion), sliding in two electromagnetic brakes fixed on the splint. When one brake was off, the bar slid freely and the corresponding dof was unlocked. When the brake was on, the bar stopped moving and the exoskeleton (by extension, the wrist) was fixed at a given joint angle, transferring forces related to the dofs directly from the hand to the forearm, bypassing the wrist. In this prototype, both brakes were activated simultaneously in on/off mode, with the person pressing a foot-operated pedal at the time and in the joint position of their choice.

Three experimental conditions were compared: (1) without exoskeleton (*NoExo*), (2) wearing the exoskeleton with brakes off i.e., passive (*PsfExo*) and (3) using it with brakes on i.e., active (*ActExo*). *NoExo* was the reference condition for the analysis, *PsfExo* primarily aimed at detecting the risks associated with wearing the exoskeleton, and *ActExo* aimed at quantifying the potential benefits of using it. The order of experimental conditions was randomized into balanced groups. At the beginning of each experimental condition, a newly sharpened knife was provided.

Measurement consisted in recording cutting forces, motion capture and surface EMG data. The cutting forces were recorded by a 6-dof sensor (HBK MCS10-005-6C) positioned under the cutting table. Motion capture was performed using an optoelectronic system with 22 cameras (Qualisys®, 200Hz): to assess the right upper limb and trunk motion, participants were pre-equipped with 21 reflective skin markers following a standardized procedure [28]. Three markers (head of the third metacarpal, radius and ulna styloid processes) were repositioned after putting on and taking off the exoskeleton; their position was marked with a hypoallergenic pencil before the experiment to ensure reproducibility of their positioning. Markers facing the anterior iliac spines and xiphoid process were placed on the chain mail apron (Fig. 1.A). EMG activities were recorded using surface electrodes (Delsys Trigno®, 2000Hz) at the ulnar forearm flexor bundle (i.e., flexor carpi ulnaris and palmaris longus muscles), forearm extensor bundle (extensor carpi ulnaris muscle), biceps brachii, long head of the triceps brachii, anterior, middle, and posterior deltoid, upper trapezius, latissimus dorsi, right and left erector spinae muscles at the level of the first lumbar vertebrae, following SENIAM guidelines [29].

After familiarizing themselves with the tasks and the exoskeleton, participants performed a task mobilizing all the joints of their upper limbs and trunk, for future use to geometrically calibrate the osteo-articular model (see below). Then, participants performed the range of motion (rom) task consisting of a maximum pronosupination movement with the elbow flexed at 90°, at 90° of shoulder abduction in shoulder external rotation, then a maximum wrist flexion-extension and a maximum radio-ulnar deviation with the elbow flexed at 90° and with slight abduction of the shoulder. This rom task was carried out 6 times in the *NoExo* and *PsfExo* conditions. Participants then performed three standardized foam-cutting tasks corresponding to a variety of gestures that we identified as commonly performed in actual meat cutting tasks, based on observations and discussions in a plant (Fig. 1.A): straight cut drawn on the top of the foam with the knife gripped like a dagger, straight cut drawn on the side of the foam and 100mm diameter semicircular cut drawn on the front of the foam, both with a traditional grip. Preliminary tests showed that the d=24 kg/m³ density polyurethane foam used in the experiment could achieve a level of cutting forces similar to that encountered in actual meat cutting tasks [11], [30]. Tasks began in the anatomical position. Participants then knocked the knife on the table, indicating the beginning of the reaching phase. They were free to position themselves as they wished with respect to the table. The tasks were performed with the knife in the right hand, without the left hand touching the table. Each task was performed six times in each experimental condition (Table I). The exception was the semicircular task, which could not be performed with the active exoskeleton and was therefore performed only in the *NoExo* and *PsfExo* conditions. Tasks were performed in random order, in sets of 3 repetitions of the same task, with balanced groups.

The foams were collected to analyze the cutting accuracy. Participants also completed a user experience questionnaire (French and English versions in supplementary material). This questionnaire was inspired by the QUEST 2.0 (Quebec User Evaluation of Satisfaction with Assistive Technology), a generic scale for assessing satisfaction with assistive technology, which is customizable for specific applications [32]. Biomechanists, one medical doctor, one butchery expert and exoskeleton designers worked together to ensure that the questionnaire covered all important aspects of the exoskeleton prototype evaluation. Complementarity was sought with instrumented measurements. The dimensions covered were the ease and comfort in performing tasks, the comfort of the exoskeleton and the relationship between foam-cutting tasks and real-life tasks. In addition to this main questionnaire, after each repetition of each task performed in the *ActExo* condition, participants completed a questionnaire assessing how they have been able to use the exoskeleton; their degree of satisfaction with the locked position and their ability to rest on the exoskeleton were assessed. Questionnaires were self-administered, based on analog visual scales with additional possibility of verbal comments.

TABLE I
EXPERIMENTAL DESIGN. *NoExo*: NO EXOSKELETON.
*PsfExo*: PASSIVE EXOSKELETON. *ActExo*: ACTIVE
EXOSKELETON *AQ*: QUESTIONNAIRE ASSESSING THE ABILITY TO
USE THE EXOSKELETON

| | | Experimental conditions (random order) | | |
|---|---|---|---|---|
| | | NoExo | PsfExo | ActExo |
| Calibration trial | | 1 | | |
| Wrist range of motion task | | 6 | 6 | |
| Cutting tasks Order randomized by groups of 3 repetitions | Top | 6 | 6 | 6 (task + AQ) |
| | Lateral | 6 | 6 | 6 (task + AQ) |
| | Semicircular | 6 | 6 | |
| Comfort questionnaire | | 1 | 1 | 1 |

*B. Data processing*

The cutting task was defined as the foam cutting phase (force magnitude >1N). The reaching phase took place between the impact of the knife on the table and the beginning of the task (Fig.1.B).

The operational impact of wearing or using the exoskeleton was assessed in terms of reaching phase and cutting task durations (s) and cutting inaccuracy, which was defined as the area between the trajectories drawn on the foam and the actual cutting path. This surface was colored (Fig. 1.C), scanned at 300ppp and filtered using the colorThresholder application of Matlab to obtain a black and white image. After visual inspection of the segmentation quality, the surface of the black area was calculated from the number of black pixels. The average cutting force during the task (N) was also computed.

The impact of the exoskeleton on joint angles was assessed based on maximum wrist angles during the wrist rom task and on maximum upper limb and trunk joint angles during the cutting tasks. An osteo-articular model of the trunk (2 segments, 6 dof) and upper limb with clavicle (4 segments, 9 dof) [33] was implemented in the opensource library CusToM, which supports the classic motion analysis functions (kinematics and inverse dynamics computation) [34]. Marker trajectories were low pass filtered with a no phase shift 10Hz 4th order Butterworth filter. Geometrical calibration of the model was performed from the experimental (marker) data, using an optimization routine minimizing the error by changing the segment length [35]. Joint angles were computed using the international society of biomechanics recommendations [36], except for the sternoclavicular joint, for which the XZ angle sequence was used. To obtain an overview of the trunk angles, the lower and upper trunk angles were summed axis by axis. Due to the small angles on the first two axes of the Euler angles (<7° in each experimental condition), this approximation was assumed to be acceptable.

The impact on user effort was assessed by inverse dynamics and EMG activity. Body segment mass and inertial parameters (BSIP) were first scaled to the participants' height and mass using standard anthropometric BSIP predictive models [37] in CusToM. Then, the mass of the equipment was integrated into this model as point masses assigned to the corresponding limb segments at the center of mass of the equipment (Fig. 1.A). Cutting forces were applied to the hand of the subject. Finally, joint torques were computed using a Newton-Euler recursive algorithm implemented in CusToM, with joint angles, their derivatives and external forces as input data [38]. Joint torques were then recalculated under the assumption that the exoskeleton was weightless, to distinguish between overload caused by kinematic adaptation to the presence of the exoskeleton or due specifically to its mass. It should be noted that wrist torque computation only provides resultant torques which cannot be isolated from each other, i.e., the sum of participants' joint torques and force transferred via the exoskeleton in the *PsfExo* and *ActExo* conditions.

After visually checking the quality of EMG, the raw data were detrended to prevent baseline offset, then the RMS-rectified and smoothed EMG signals were calculated using a sliding window of 100ms [39]. Mean EMG activities were computed across time during the reaching and the task, for each trial, and then normalized by dividing them by their mean values for each person in the *NoExo* condition [40].

*C. Statistics*

Descriptive statistics were carried out using R [41]. Then,

linear mixed-effects models with a random subject effect at origin and fixed effects were established using the *lmer* function of R [42]. First, the effect of the tasks and repetition number were successively tested as fixed effects to give an overview of their effect on the dependent variables (Table II). Due to the expected effect of the task and the fact that the semicircular cutting task was not performed in the *ActExo* condition, the effect of the exoskeleton was not assessed at this stage, but only in multivariate models performed on a task-by-task basis, to avoid confusing the effect of the task with the effect of the exoskeleton itself. Fixed effects in these multivariate models encompassed the experimental conditions, the number of repetitions and, for explaining joint torques and EMG activities during tasks, the mean cutting force. They were simplified using a top-down procedure. In the event of statistically significant effects of the experimental condition, post-hoc tests were carried out. The *NoExo* condition was set as a reference, so that it could be compared with the *PsfExo* condition and the *ActExo* condition. The model residuals were checked for normality. To meet this condition of model validity, the cutting inaccuracy (mm²) was square rooted before being included in statistical models. The wrist rom task was processed separately, using the same procedure.

Regarding user experience questionnaires, the small number of statistical observations precluded the use of linear models. Paired Student's t-tests were performed. In addition, Pearson linear correlations between the comfort of the exoskeleton and potential sources of discomfort were sought.

## III. RESULTS

As expected, almost all dependent variables were task-dependent (Table II). The number of repetitions had less frequent effects, indicating relatively satisfactory familiarization with the tasks and with the exoskeleton: only the approach phase (-0.07s per occurrence, $p<0.05$) and the cutting task duration (-0.08s per occurrence, $p<0.001$) were affected, along with most wrist maximal joint angles during the rom task.

TABLE II
IMPACT OF THE TASK AND OF THE NUMBER OF REPETITIONS ON DEPENDENT VARIABLES: % OF STATISTICALLY SIGNIFICANT TESTS. R: REACHING. T: TASK. GREY CELLS: NOT APPLICABLE

| Dependent variables | Task | | Repetitions | |
|---|---|---|---|---|
| (Number of tests) | R | T | R | T |
| **Operational impact** | | | | |
| Duration (1 test) | 100 | 100 | 100 | 100 |
| Mean force (1 test) | | 100 | | 0 |
| Cutting accuracy (1 test) | | 100 | | 0 |
| **Maximum joint angles** | | | | |
| Wrist rom tasks (6 angles) | | | | 83 |
| Cutting tasks (24 angles) | 96 | 100 | 0 | 0 |
| **User effort** | | | | |
| Maximum joint torques (24 torques) | 88 | 100 | 8 | 0 |
| EMG (11 muscles) | 64 | 91 | 10 | 0 |
| **User experience** | | | | |
| Main questionnaire (8 questions) | | 0 | | |
| Use achievement (2 questions) | | 100 | | 0 |

### A. Operational impact

Using the exoskeleton locked was associated with longer reaching phases and tasks durations ($p<0.05$ to $p<0.001$) (Fig. 2). Accuracy was improved in the *PsfExo* ($p<0.001$) and *ActExo* ($p<0.001$) conditions for the top cutting task but not significantly (NS) deteriorated for the other two tasks. Cutting forces were generally NS modified. The average cutting forces were 25.3±6.6N in the top cutting task, 19.7±6.0N in the side cutting task and 14.2±3.8N in the semicircle cutting task.

### B. Impact on maximum joint angles

Wearing and using the exoskeleton had an impact on joint angles, both in terms of maximum wrist amplitude during the wrist rom task and of maximum upper limb joint angles reached during cutting tasks. The quality of the kinematic reconstruction was acceptable [35], with a mean error of 1.0cm for the wrist rom task and 1.5cm for the cutting tasks.

Regarding the wrist, maximum angles were all altered in the wrist rom task by wearing the exoskeleton (*PsfExo*), except pronation, which amplitude was preserved (Fig. 3): flexion, extension and supination were reduced, and radio-ulnar deviation amplitudes shifted towards the ulnar zone ($p<0.001$ for each angle). During reaching phases and cutting tasks, the maximum angles reached were generally much lower than those observed in wrist rom tasks (Fig. 4.A, and Fig. 1 and 2 in supplementary material). In the *NoExo* condition however, joint angles thresholds associated with an increased risk of MSDs [24] were reached in ulnar deviation (23.4±6.1° instead of <20°) and forearm pronation (69.8±5.4° instead of <60°) during the semi-circular cutting task. In this task, which involved the greatest pronosupination rom, pronation was preserved and supination reduced ($p<0.001$), as it was in the wrist rom tasks. A significant shift in radial and ulnar deviation towards the ulnar area was generally present during reaching phases and cutting tasks in the *PsfExo* and *ActExo* conditions. Reduced wrist flexion-extension amplitudes were observed during reaching phases and during the lateral cutting task, due to decreased wrist extension or flexion. In the top cutting task, a shift toward flexion was observed.

Reaching and tasks were constantly performed in glenohumeral elevation, and slight anterior and right lateral trunk flexion. In the *NoExo* condition, safety thresholds [24] were exceeded in elbow extension (48.1±10.4° instead of >60°) and glenohumeral elevation (81.4±5.6° instead of <80° of shoulder antepulsion or <45° of shoulder abduction) during the lateral cutting task. The patterns of elbow maximum flexion angle and glenohumeral joint angles modifications according to experimental conditions were unsystematic. The most remarkable feature at the trunk level was its low mobility and the small differences between experimental conditions, of less than 3.0° in all dof.

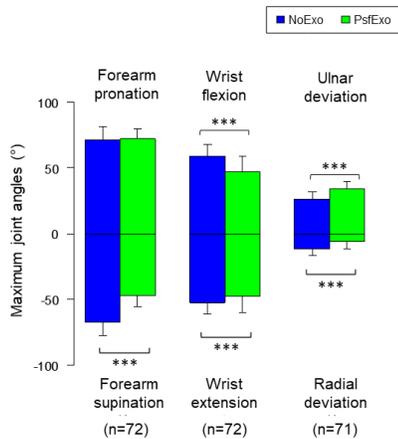

**Fig. 3.** Maximum wrist joint angles (mean, standard deviation) during the wrist range of motion task. *** $p<0.001$. *NoExo*: no exoskeleton. *PsfExo*: passive exoskeleton

*C. Impact on user effort*

Wearing and using the exoskeleton had an impact, both on joint torques and EMG. The effects of the exoskeleton on joint torques were pronounced at the shoulder level, notably on glenohumeral elevation torques. Their increase between the *NoExo* and *ActExo* conditions reached 25.1% ($p<0.001$) and 39.7% ($p<0.001$) in the top and lateral cutting tasks respectively, and 29.9% ($p<0.001$) and 40.8% ($p<0.001$) in the related reaching phases. These effects diminish or disappear with the virtually weightless exoskeleton (see for example glenohumeral elevation on Fig. 4 in supplementary material). Thus, these effects were mainly attributable to the mass of the exoskeleton, although kinematic adaptations also seemed to play a role, e.g. in increasing shoulder joint torques in the *ActExo* condition during the lateral cutting task. Wearing and using the exoskeleton increased elbow flexion torques and decreased extension torques ($p<0.001$). Maximum joint torques reached in the *NoExo* condition were -8.31±1.80Nm in glenohumeral elevation during the top cutting task and 8.59±1.95Nm in elbow flexion during the lateral cutting task.

The exoskeleton condition had non-systematic or NS effects on wrist and trunk joint torques. At the wrist, maximum radial deviation torques were observed in the lateral cutting task (-5.6±2.0Nm in the *NoExo* condition), and maximum ulnar deviation torque in the top cutting task (1.9±0.9Nm, *NoExo* condition). This pattern was expected, given the type of knife grip used, traditional and dagger grip, respectively. The maximum pronation torques were observed during the semi-circular cutting task (3.18±0.98Nm, *NoExo* condition). The trunk was loaded systematically in posterior flexion and often in left lateral flexion, which was expected given the position of the upper limb and trunk.

Compared with the *NoExo* condition, the wrist flexors' EMG activity was reduced in the *ActExo* condition during the top ($p<0.01$, Fig. 6 in supplementary material) and lateral ($p<0.001$, Fig. 5) cutting tasks, of 18.7% and 17.5%, respectively. The exoskeleton condition had non-systematic or NS effects on wrist extensors' EMG. When significantly modified, shoulder muscles' (upper trapezius, deltoid, and biceps brachii) EMG activity was increased in the *PsfExo* and *ActExo* conditions, both during the reaching phases and cutting tasks. The increase in upper trapezius' EMG activities between the *NoExo* and *ActExo* conditions reached 61.7% (NS) and 44.8% ($p<0.05$) in the top and lateral cutting tasks, respectively. On the contrary, erector spinae muscles were generally not significantly over-activated when using or wearing the exoskeleton. The wide variability in EMG activities of certain muscles, such as the upper trapezius in the *ActExo* condition, should also be noted.

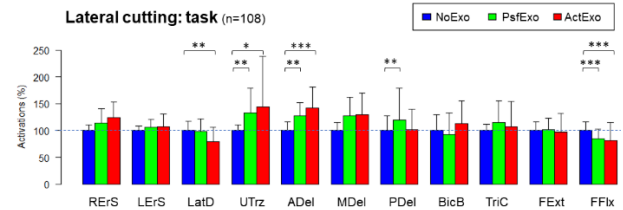

**Fig. 5.** EMG activity during the lateral cutting task (mean, standard deviation). *** $p<0.001$, ** $p<0.01$, * $p<0.05$. *NoExo*: no exoskeleton. *PsfExo*: passive exoskeleton. *ActExo*: active exoskeleton. *RErS, LErS*: right and left erector spinae muscles, respectively. *LatD*: latissimus dorsi. *UTrz*: right upper trapezius. *ADel, MDel, PDel*: right anterior, middle and posterior deltoid. *BicB*: right biceps brachii. *TriC*: right triceps brachii, long head. *FExt*: right forearm extensor bundle. *FFlx*: right forearm flexors bundle

*D. Impact on user experience*

Wearing and using the exoskeleton had NS effect on user experience (Fig. 6). However, the statistical power of the statistical tests was low, as the risk of omitting a strong effect - according to the Cohen classification [43] - was $\beta=65\%$ when the number of statistical observations was $n=18$. The exoskeleton was evaluated as quite comfortable. Mobility of the exoskeleton in relation to the skin ($p<0.01$), dissatisfaction with exoskeleton dimensions (NS) and the need to grip the knife handle tighter (NS) had strong [43] pejorative effects on exoskeleton comfort ($r>0.7$). The top and lateral cutting tasks were rated as realistic, while the semi-circular cutting task was rated as averagely realistic; open-ended comments highlighted the smaller diameter of the real deboning tasks and the highest difficulty of executing circular cutting movements in foam, despite the selection of this larger diameter. Participants were moderately enthusiastic about working with this exoskeleton in the coming months, although there were wide interpersonal differences in appreciations (Fig. 6). Satisfaction with the locked position was good (8.1±1.5/10 for the top cutting task and 7.2±2.0/10 for the side cutting task), as was operators' self-assessment of their ability to rest on the exoskeleton (7.8±1.7/10 and 6.8±2.1/10, respectively).

## IV. DISCUSSION

The aim of this study was to assess the potential effectiveness and risk of a prototype of wrist exoskeleton. Simplified, reproducible tasks designed to be similar to meat cutting tasks were performed. As expected, the average cutting forces were realistic, ranging from 14.2 to 25.3N depending on the task, compared with 19N in real in situ pork shoulder deboning tasks [11], [30]. The choice of cutting trajectories, corresponding to a selection of meat cutting gestures based on field observation and discussions with an expert in butchery, was globally validated by the butchers involved in the study, who rated the tasks as realistic overall.

*A. Potential effectiveness and risk analysis*

Based on previous epidemiological [9], [10] and biomechanical [11] studies, a reduced load was desired, particularly in wrist flexion. Given the impossibility of distinguishing between joint torques and torques transferred by the exoskeleton, the relevant criterion to assess the load in wrist flexion was the level of wrist flexors' EMG activity. The decrease in wrist flexors' EMG activity during use of the active exoskeleton, of up to 18.7%, does not seem negligible from a clinical point of view. Literature data are consistent with this result. With an active wrist exoskeleton tested on four young men carrying 1.5 kg statically or during wrist flexion from 0 to 60° at fixed speed, Chiaradia et al. [7] obtained effects of the same order of magnitude on wrist flexors, -15.6% to -24.1% depending on muscle groups and tasks. These results legitimize further evaluation of the prototype's effect on MSDs risk factors or on the onset of MSDs in people performing meat cutting tasks.

This work showed no major operational impact of wearing or using this exoskeleton, except for increased durations of reaching phases and cutting tasks while using the active exoskeleton. These increased durations can probably be explained by the need to lock the exoskeleton with a control pedal, which represents an additional task, most often performed during the reaching than during the cutting tasks, according to our qualitative observations during the experiment. This phenomenon was mitigated by the number of repetitions (Table II), indicating incomplete familiarization of participants with the exoskeleton. Additionally, as no effect on user experience were found, it could be concluded that, overall, no major usability issues were identified for this prototype.

The study also aimed to anticipate the risk of physical overload associated with the use of the exoskeleton. In the event of overloading, we wanted to distinguish between overloads due to kinematic adaptation caused by the presence of the exoskeleton and overloads due specifically to its mass, in order to make recommendations on the design of the exoskeleton. In our experimentation, EMG data showed that wearing the exoskeleton mainly overloaded the shoulder region. Since shoulder tendinitis are already common in the meat processing industry [10], particular attention should be paid to this risk during subsequent phases of the exoskeleton's development. Furthermore, as the estimation of joint torques computed with and without the exoskeleton mass showed that this overload was mainly attributable to the mass of the exoskeleton (670g, located relatively far from the forearm longitudinal axis), it is important that portable wrist exoskeletons remain relatively light, with a mass distribution that does not create large moments [13] that may fatigue the user unnecessarily. To achieve this goal, several possibilities can be imagined, ranging from exoskeleton proximal weight transfer systems to assistance systems combining active actuators, which are often heavy, and lightweight passive assistance, depending on the dof. Regarding the massive increase in EMG activity of the upper trapezius, other phenomena such as co-contractions due to incomplete familiarization or measurement artifacts could be concomitantly at stake [44], [45].

*B. Study limitations*

Firstly, a small number of tasks, or rather of simple subtasks, were explored. Therefore, our study does not account for all meat cutting gestures. This is an important limitation, as significant differences among tasks were found for all parameters explored. It has been shown in the literature that meat cutting tasks include gestures reaching the angles of radio-ulnar deviation and wrist extension [11] that we observed in the wrist rom task. These kinematically demanding cutting subtasks were not explored in our study. Nevertheless, the tasks we chose were demanding, as they often exceed the safety thresholds for joint angles [24]. Secondly, the context in which tasks are performed can modify results. Laboratory conditions differ from ecological conditions, with less time pressure, fewer repetitions of tasks than in a day's work and without the cold, noisy environment of meat cutting plants [46]. Moreover, the way operators perform tasks can be modified by measuring equipment. Finally, the number of participants was low, to avoid over-dimensioning a study designed to provide an early assessment of the prototype. Beyond the possible variability of effects due to sampling fluctuations, it should be noted that the assessment of user experience had particularly limited statistical power since the questions were asked only once per exoskeleton condition; the risk of missing relevant effects on user experience is therefore substantial. Consequently, the results obtained in our study are largely preliminary.

Despite these limitations, laboratory evaluation enables precise measurement of a wide range of parameters [2] to assess the impact of an exoskeleton, with advanced equipment and controlled tasks [47]. Motion analysis using an optoelectronic system and external force measurements are indeed the gold standard for estimating joint angles and joint torques. The results may still be affected by various sources of error, for example errors in kinematic reconstruction propagated to the analysis of joint torques [48], and inaccuracy in estimating the position of the equipment mass on the different body segments. Additionally, EMG provides a direct, global estimate of the activity of selected muscles. Overall, the greatest risk of misjudging the benefits and risks associated with using this exoskeleton lies in the fact that the criteria studied were intermediate criteria. These criteria enable a quick and analytical estimation of expected effects, which is appropriate in the early phase of exoskeleton development. However, fatigue studies performed in the field and, even more so, the evaluation of the actual medical effect i.e., the actual onset of MSDs [49], still remain to be conducted with a more advanced version of the prototype, alongside a study of acceptance of the exoskeleton in the workplace.

V. CONCLUSION

This study proposes a method for evaluating an exoskeleton prototype, which includes the analysis of its operational, joint angles and muscular impacts, as well as its impact on user experience. Despite the small number of participants, the reduction in wrist flexor muscles' activity indicates that positive effects on the risk of MSDs are possible. The analysis also suggests the existence of risks in the future deployment of this exoskeleton in the workplace, particularly at the shoulder level. Exoskeleton design elements requiring attention to mitigate this risk were identified, namely the mass of the exoskeleton and its distribution with respect to the segment. The proposed multi-criteria evaluation has provided guidance for the next stages of prototype design and evaluation, especially to ensure operator safety.

ACKNOWLEDGMENT

The authors thank COOPERL company[1] workers for their participation, Alain Lucas for his expertise in butchery and the LAB4i company[2] for their support during the experimentation.

---

[1] https://www.cooperl.com/

[2] https://groupe-ovalt.com/lab4i/

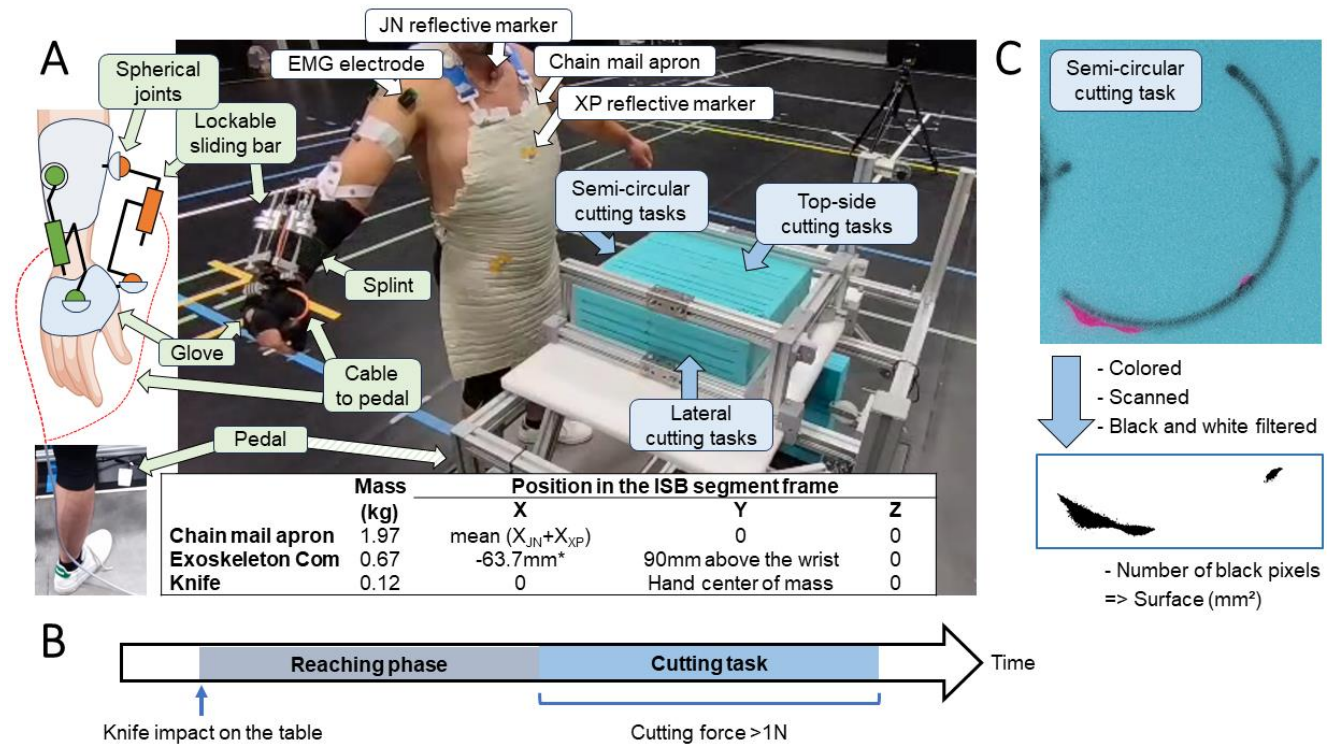

**Fig. 1. A.** Participant equipment (white labels), exoskeleton (green labels) and drawing of the cutting tasks on the foam (blue labels). **B.** Timeline of one task. **C.** Processing of cutting inaccuracy for a semi-circular cutting task. *ISB*: international society of biomechanics [36]. *Com:* center of mass of the exoskeleton, located between its brakes. *JN*: jugular notch. *XP*: xiphoid process. *63.7mm = distance measured between the exoskeleton center of mass and the skin surface (25 mm) + average maximum radius of the forearm in men in the literature (38.7mm, [31]).

...

> REPLACE THIS LINE WITH YOUR MANUSCRIPT ID NUMBER (DOUBLE-CLICK HERE TO EDIT) <

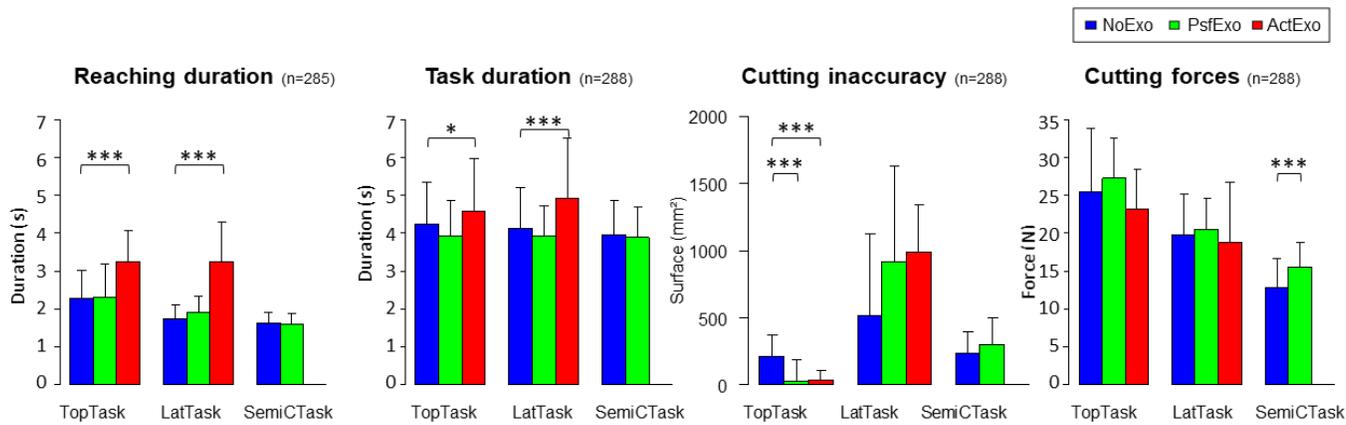

**Fig. 2.** Operational impact of wearing or using the exoskeleton during cutting tasks (mean, standard deviation). *** p<0.001, ** p<0.01, * p<0.05. *NoExo*: no exoskeleton. *PsfExo*: passive exoskeleton. *ActExo*: active exoskeleton. *TopTask*: top cutting task. *LatTask*: lateral cutting task. *SemiCTask*: semi-circular cutting task.

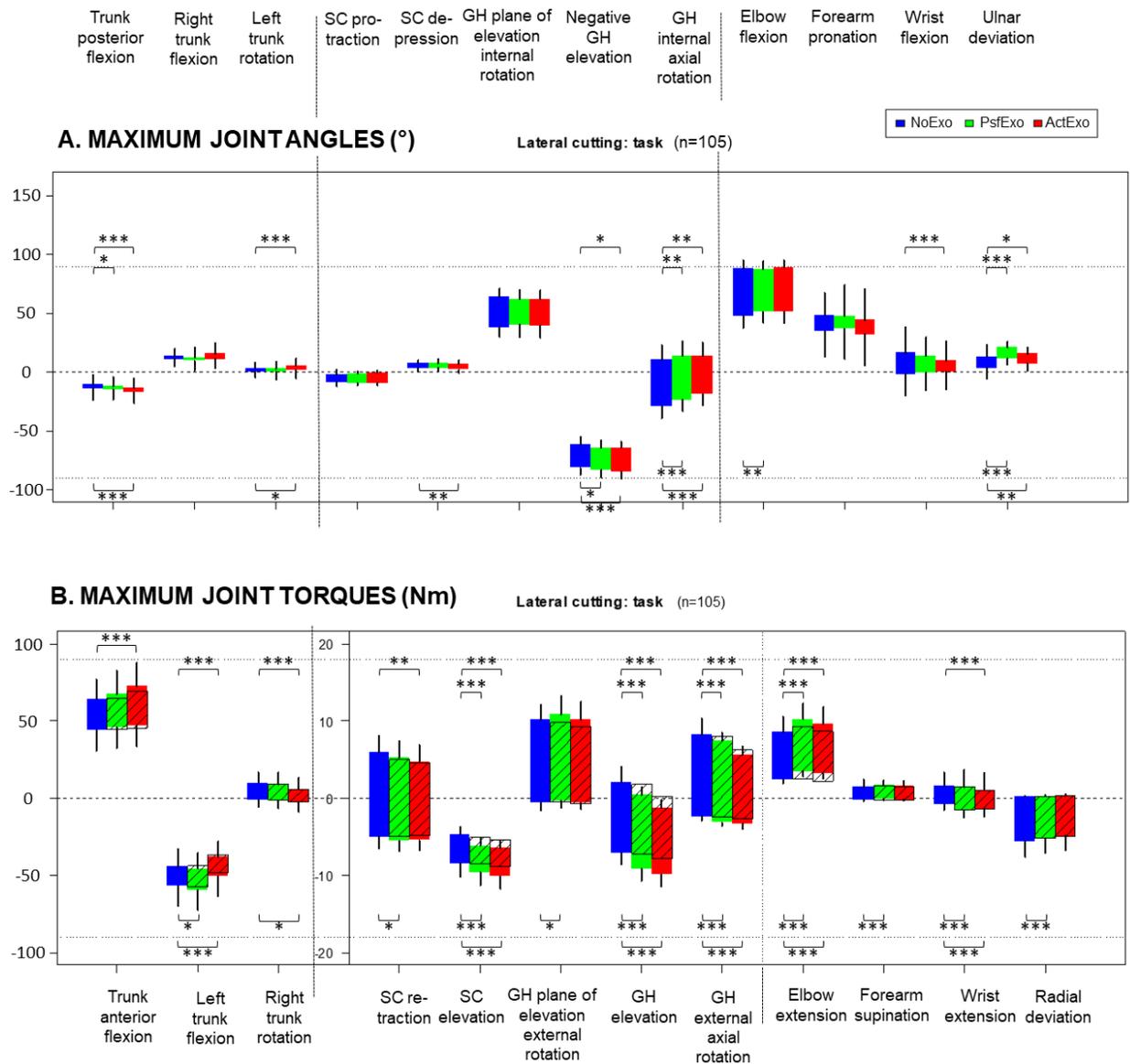

**Fig. 4. A.** Maximum joint angles during the lateral cutting task (mean, standard deviation). **B.** Maximum joint torques (mean, standard deviation). Hatched areas: simulated joint torques with a virtually weightless exoskeleton. *** p<0.001, ** p<0.01,

* p<0.05. *NoExo*: no exoskeleton. *PsfExo*: passive exoskeleton. *ActExo*: active exoskeleton. *GH*: glenohumeral joint. *SC*: sternoclavicular joint.

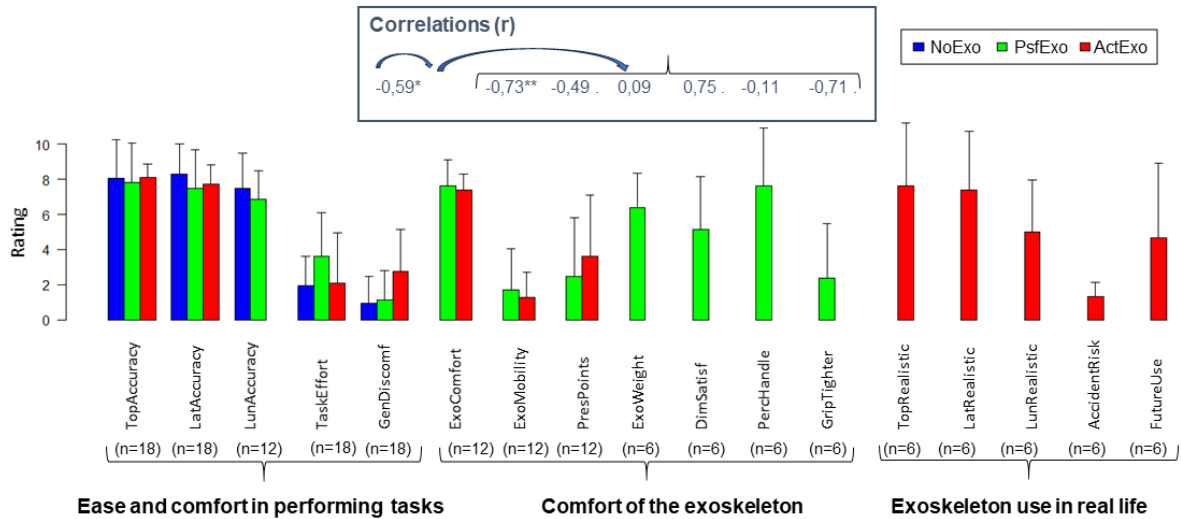

**Fig. 6.** User experience: rating out of 10 (mean, standard deviation). Legends, top to bottom and left to right: * p<0.05, ** p<0.01. *NoExo*: no exoskeleton. *PsfExo*: passive exoskeleton. *ActExo*: active exoskeleton. *TopAccuracy*, *LatAccuracy*, *LunAccuracy*: ease in performing top, lateral and semi-circular cutting tasks with precision. *TaskEffort*, *GenDiscomfort*: effort and general comfort in performing tasks. *ExoComfort*: exoskeleton comfort. *ExoMobility*: mobility of the exoskeleton in relation to the skin. *PresPoints*: pressure points. *ExoWeight*: participant satisfaction with exoskeleton weight. *DimSatif*: participant satisfaction with exoskeleton dimensions. *PercHandle*: ease to feel the knife handle. *GripTighter:* need to squeeze the hand more to hold the knife. *TopRealistic*, *LatRealistic*, *LunRealistic*: task realism. *AccidenRisk*: perceived risk of accident. *FuturUse*: feeling about future use.